**This is a revised version of the preprint, which is of an article that is under review and has received feedbacks from reviewers, as of 10 Feb 2023.**

**We've made several important modifications to the paper, including updating the title, completely revising the discussion section (now Section 6), and making various other minor adjustments throughout the paper.**

# Title Page

**Title**
Supporting data discovery: A meta-synthesis comparing perspectives of support specialists and researchers

**Names of the authors**
Guangyuan Sun, Tanja Friedrich, Kathleen Gregory, Brigitte Mathiak

**Affiliations, addresses, and ORCID of the authors**
- Guangyuan Sun, National Institute of Education, Nanyang Technological University, Singapore. https://orcid.org/0000-0001-7352-2158
- Tanja Friedrich, German Aerospace Center, Scientific Information, Cologne, Germany. https://orcid.org/0000-0003-1557-3728
- Kathleen Gregory, University of Vienna, Vienna, Austria; School of Information Studies and Scholarly Communications Lab, University of Ottawa, Ottawa, Canada. https://orcid.org/0000-0001-5475-8632
- Brigitte Mathiak, GESIS - Leibniz Institute for the Social Sciences, Cologne, Germany. https://orcid.org/0000-0003-1793-9615

**E-mail address of the corresponding author**
Guangyuan Sun, gsun003@e.ntu.edu.sg

**Abstract**

*Purpose* Data discovery practices currently tend to be studied from the perspective of researchers or the perspective of support specialists. This separation is problematic, as it becomes easy for support specialists to build infrastructures and services based on perceptions of researchers' practices, rather than the practices themselves. This paper brings together and analyzes both perspectives to support the building of effective infrastructures and services for data discovery.
*Methods* This is a meta-synthesis of work the authors have conducted over the last six years investigating the data discovery practices of researchers from different disciplines, with a focus on the social sciences, and support specialists. We bring together and re-analyze data collected from in-depth interview studies with 6 support specialists in the field of social science in Germany, with 21 social scientists in Singapore, an interview with 10 researchers and 3 support specialists from multiple disciplines, a global survey with 1630 researchers and 47 support specialists from multiple disciplines, an observational study with 12 researchers from the field of social science and a use case analysis of 25 support specialists from multiple disciplines.
*Results* We found that there are many similarities in what researchers and support specialists want and think about data discovery, both in social sciences and in other disciplines. There are, however, some differences which we have identified, most notably the interconnection of data discovery with web search, literature search and social networks.

*Conclusion* We conclude by proposing recommendations for how different types of support work can address these points of difference to better support researchers' data discovery practices.

 **Keywords**

**Statements and Declarations**

The authors have no competing interests to declare that are relevant to the content of this article.

**Author contributions:**

All authors contributed to the study conception and design. All authors contributed the methods and findings of their respective studies. The first draft and consecutive reviews were done collaboratively by all authors. All authors contributed equally to the discussion and conclusion. Literature review was primarily done by Guangyuan Sun. Guangyuan Sun also developed the typology of support work.

# 1. Introduction

Research data management support is an increasingly visible component in the host of support services provided by universities and research institutions. While data librarians, archivists and educational specialists provide support with managing and sharing data [1–4], they are also interested in supporting practices afforded by well-managed data, such as data discovery.

Data discovery practices currently tend to be studied from the *perspective of researchers* or *the perspective of support specialists*. While work from both perspectives aims to inform the design of tools, infrastructures and services, these perspectives often exist separately from each other in the literature.

Studies from the perspective of researchers and data seekers situate data discovery and reuse in relation to disciplinary norms or types of data [5–8]; the use of particular repositories or portals [9, 10]; or the influence of trust, reputation and ease of access on the process [11–13].

Work conducted from the perspective of data support specialists focuses on typical workflows [3, 14, 15], tasks [16, 17] and the details of infrastructural projects (e.g., the FREYA project[1], or the GeRDI project[2]) which are designed to support data discovery and reuse.

With few exceptions (i.e., [1]), these two perspectives exist separately from each other in the literature. This separation is problematic, as it creates an environment where it becomes easy for support specialists to build infrastructures and services based on perceptions of researchers' practices, rather than the practices themselves. We argue that we need to bring together and analyze both the practices and perspectives of researchers and those of data support specialists in order to build effective, sustainable infrastructures and services to support data discovery.

This paper brings together and re-analyzes data from past work, rooted in both perspectives, in a systematic meta-synthesis. We draw on six of our own studies, conducted over a six-year period, to closely examine differences and points of alignment between the data discovery practices of researchers and the perspectives and existing work of support specialists.

We begin by reviewing the literature to develop a typology of the different types of work done by support specialists. We then re-analyze and synthesize the data from our past studies, according to both the perspective of researchers and support specialists and look for points of alignment between them. These studies include data from in-depth interview studies [6, 18, 19]; a global survey [20, 21]; an observational study [22] and a use case analysis conducted by information professionals [23, 24]. We conclude by relating five emergent themes from our analysis to our typology of support work, highlighting both existing best practices and identifying areas for improvement.

# 2. Literature Analysis

The aim of this study is to compare the divergent perspectives of researchers and support specialists with regard to data discovery. We provide a succinct overview of literature from the perspective of researchers (Section 2.1) but focus on the perspective of support specialists (Section 2.2) in order to develop a typology of support specialist work which we use in our analysis.

---

[1] https://www.project-freya.eu/en
[2] https://www.gerdi-project.eu/

## 2.1 Prior studies from the perspective of researchers

Many data discovery studies are motivated by studying researchers' data reuse behaviors [7, 19, 25, 26]. They correlate concepts, such as attitude, motivation and disciplinary norms in data reuse with discovery practices, e.g., through surveys.

Qualitative studies, such as [27] investigate workflows of data search in a web-based environment and introduce patterns of web-based data search, i.e. discover, explore and analyze. They characterize data discovery as the "serendipitous discovery of datasets…by means of descriptive metadata", where the input is "unstructured free text" and the output is "list of (relevant) datasets". Koesten et al. [28] categorize users' data-centric tasks into two types: *process-oriented* (e.g., using data for machine learning processes) or *goal-oriented* (e.g., using data to answer a question), and identify different user information needs for the different types. Wu et al. [29] studied user requirements for users of STEM data repositories and collected nine user requirements. Bishop et al. [30] studied the information seeking behavior of 22 environmental scientists with regards to data reuse. They found many scientists performed known-item searches in trusted sources, and that half began their search with a web search on specific keywords or names of the data repository.

Search query and log analysis are a third way to investigate data search behavior e.g., of structured data stored on the web [28], in open government data portals [31] or in research data repositories specifically [32–34]. They yield interesting results, such that data search queries are typically brief and that users tend to discover data repositories through search engines such as Google. However, many of these studies are limited in scope, focusing specifically on the source of log analysis and the results are restricted to the structure of the repository and the types of data stored.

## 2.2 Types of work conducted by support specialists

Studies from the perspective of support specialists regarding discoverability of data are comparably sparse. We found that there is a body of research to be compiled from studies or descriptions of professionals with different roles in research data management support. There are many terms used to describe research data support staff such as data curator, data librarian, data manager, data steward, data specialist and research information scientist [35]. There is a lack of common terminology and a variability of both position titles and actual duties. Research data support work, as we choose to call it, can be both social and technical, involving technical knowledge on research data management (RDM) systems and social knowledge for RDM community building [36].

We identify three primary types of support work from the literature: people-oriented roles, e.g., providing consultations (Section 2.2.1); metadata-related roles, e.g., providing data documentation services which are non-technical but focus on enhancing the completeness of research data description and data discoverability (Section 2.2.2); and infrastructure-oriented roles, e.g., providing IT-focused technical tasks (Section 2.2.3). It is important to note that these types of work are not mutually exclusive, as support specialists may be required to fulfill multiple roles depending on their job requirements. Nonetheless, the proposed categorization serves as a heuristic device for structuring and discussing the complexities of RDM support work.

## 2.2.1 People-oriented support work

People-oriented research support work varies between organizations. It depends on technical infrastructure, expertise of staff, and the organization's value, culture and structure [37]. At the libraries of large universities, such support work includes the topics of research data management, open access, scholarly publishing, research impact measurement, developing research guides, and providing research consultations and research tools recommendations [38]. Some of this work can be seen as an extension of traditional library services to research data [3]. Tammaro et al. [35] further classify skills of data curators into four categories: *data management* (e.g., to advise on file naming conventions), *data description and documentation* (e.g., to provide consultation on metadata creation following standardized schemas), *data deposit/publishing* (e.g., to help with data anonymization) and *archiving and preservation*.

Cox et al. [39] found that, across countries, libraries have provided leadership in RDM, but there are also alternative organizational models. While research support staff are often recruited from existing library staff, non-library professionals play an important role as well to serve specific communities [35]. Data stewards, e.g., are university employees, sometimes library staff, who generally hold a PhD degree in a research domain and are appointed to a school or department to engage researchers with in-depth discussions on RDM [40]. Data champions are volunteer domain researchers that are already embedded within research units, who are equipped with necessary research data knowledge and advocate good RDM [41].

In [39], the authors distinguish between *advisory services* and *technical services*, and find the former, which we consider to be people-oriented work, to be more wide-spread than the latter. The main types of advisory services include providing advice on data management plans, RDM training, maintaining a website of local RDM advice and useful resources, and providing advice on copyright. The study finds that such services were extensively developed across libraries. *Advisory services* also include providing advice regarding the "awareness and retrieval of reusable data sources"; this type of service was evaluated as being "patchy", as it was not among the most common advisory services in the investigated libraries. Similarly, a more recent study also finds that library support staff tend to place more value on consultation services, as some believe such services are more valuable to researchers than infrastructure-oriented or metadata-related support [42].

People-oriented support work is dynamic, including guiding researchers in describing their datasets with adequate metadata [43] (see also Section 2.2.2), personalizing comprehensive and executable data management workflows [44], and providing services related to text and data mining and artificial intelligence [3]. Das and Banerjee [45] observe that the support services provided by libraries have subtle changes on technological basis over time. Zakaria [46] discusses data visualization services in top university libraries. North America and Europe have more advanced people-oriented research data services by academic libraries [3], while other parts of the world are still in development and focusing on making policy, raising awareness and conceiving RDM IT infrastructure [47, 48].

## 2.2.2 Metadata-related Support Work

Rich metadata is relevant for the discoverability of both large and small datasets. Information professionals, including research data support specialists, have formed and are active members of metadata-related working groups to develop and promote standardized documentations to enhance data discovery and harvesting on the web, especially when data can be connected to other materials and resources that are also published online.

Such groups include the Controlled Vocabularies Working Group[3] and the XKOS - Extended Knowledge Organization System[4] under the Data Documentation Initiatives (DDI)[5] or specialized groups within the Research Data Alliance[6], DataCite[7] or the GO FAIR Implementation Network[8].

Within these groups, various knowledge representation techniques have been developed and subsequently employed by repositories. Both generic and domain-specific metadata standards have been developed. DataCite[9] is a generic metadata standard promoting research data citation and has been widely adopted by data archives for imputing metadata about datasets, as well as by infrastructures harvesting metadata from those archives.

The more recent techniques of Linked Open Data and knowledge graphs are also under exploration, though they have not yet been implemented in data repositories on a large-scale. Using these techniques, research data can be linked to research papers for better discoverability. The OpenAIRE Research Graph is an initiative that seeks to connect research data from a wide range of sources by utilizing common identifiers and vocabularies, thus enabling easy discoverability and reusability of the data for researchers [49]. Similarly, the Open Knowledge Research Graph promotes the interconnectedness of research papers with other research artifacts, such as datasets, thereby stimulating data exploration and discovery [50]. Its primary impact may be the interlinking of research articles, shifting the way knowledge is exchanged from a document-centric approach to a knowledge-based flow of information. Burton et al. [51] develop the *Scholix* (Scholarly Link eXchange) Framework with the goal of establishing a harmonized system for gathering and disseminating links between research articles and supporting data.

Research data can also be linked to other types of resources. Aryani et al. [52] created a *Research Graph* database of identity and some basic descriptive information of research datasets, of researchers (e.g., ORCID), of publications (e.g., title, DOI) and of research grants (e.g., title, participant list) to support linking research datasets stored in different data repositories. Wang et al. [53] explored the conversion of records in the same *Research Graph* to Linked Open Data format to enhance their accessibility to third-party web services over the Internet, thus indirectly increasing research data discoverability. They use *schema.org* for the conversion, which is a widely adopted standard for marking up structured data in web pages and promoted by major search engines such as Google [54]. In fact, the library world has started using *schema.org* to make metadata of books and papers in library catalogs more discoverable on the web as early as 2012 [55].

Despite the numerous activities listed above, many *libraries* do not provide *technical services* for "creating/transforming metadata for data", which would require knowledge and skills in standardized metadata schemata [39].

---

[3] https://ddi-alliance.atlassian.net/wiki/spaces/DDI4/pages/39911435/Controlled+Vocabularies+Working+Group
[4] https://ddi-alliance.atlassian.net/wiki/spaces/DDI4/pages/826408976/XKOS+-+Extended+Knowledge+Organization+System
[5] https://ddialliance.org/
[6] https://www.rd-alliance.org/
[7] https://datacite.org/
[8] https://www.go-fair.org/implementation-networks/
[9] https://schema.datacite.org/

### 2.2.3 Infrastructure-oriented support work

Research data management inevitably involves building and maintaining digital infrastructure that caters to user needs across the research data lifecycle. Such work, which is more technical in nature, is typically performed by support staff who are IT-oriented.

There are a few studies on the characteristics of research data infrastructure and related support work [17, 56], collecting user requirements for these infrastructures [57] and developing RDM related tools or software [58, 59]. Typically, these studies make recommendations on what infrastructure-oriented type of support work should entail.

Other infrastructure-oriented work, which also has overlaps with more metadata-oriented tasks, includes the development of federated research data repositories to support cross-disciplinary discovery of research data [60]; the design and implementation of discovery tools, such as DataCite Commons, that allow simple searches by works, individuals and organizations [61]; and the creation of data catalogs to support the discovery of sensitive research data that requires special attention paid to legal, ethical, or authorial constraints [62, 63].

Khan et al. [17] found that the major concern across all repositories was long-term maintenance of repository infrastructure and a lack of engagement from the users. Discovery of data was rated as a future need instead of a top priority. A similar finding by Smit [64] is that technical service providers' main goal is to focus on the maintenance of research data infrastructure and to assure its long-term availability to support data sharing and storage.

Infrastructure-oriented support work may need to invest more effort in catering to the specific needs of research data and improving data discoverability. Bugaje and Chowdhury [56] evaluated different types of research data repositories and point out that existing data discovery infrastructures are "makeshift adaptations" of the classic information retrieval systems. Another stream of work that needs more attention may be to design different data infrastructures based on peculiarities of different data formats. Hedeland [65] focuses on infrastructure requirements for hosting audio-visual linguistic research data, and identifies a gap between what is needed in the process of digitization (i.e. domain-specific research methods) and what is provided (i.e. approaches of generic research data management). He posits that the responsibilities of technical support staff should include closing this gap.

To enhance the existing national and disciplinary data infrastructures and services offered by data centers and digital libraries, information professionals are constructing national and international infrastructures (e.g., European Open Science Cloud [66]) to offer a streamlined method of discovering and accessing research data and services across various domains, thus simplifying the process of discovery, access and utilization of data for researchers from different fields. Smit [64] uncovered, however, that the software underlying research data infrastructure is often created and maintained by support staff who possess minimal formal education in software engineering or by organizations that do not have software development as a primary function.

### 2.3 Summary

Most RDM service providers tend to focus on people-oriented research support work like RDM advisory services and training for users. Advice on discovering reusable data on the Web are provided only by some libraries. The metadata-focused support work is, in most cases, an intellectual collaboration of support staff with other information professionals to develop standardized documentation aiming to improve the discoverability of data on the Web by both humans and machines. The infrastructure-oriented support work currently focuses on ensuring the robustness and sustainability of the infrastructure, and sometimes lacks engagement with and incorporation of the need of end users of the service (i.e., researchers).

There is also a gap in research investigating the actual support work provided and the services needed by users. In order to address this issue, we need to understand perspectives from both researchers and support staff. The rest of this paper compares the differences in perspectives by doing a meta-synthesis of work we have conducted in the past six years to draw insights that may inform the future development of support work for data discovery.

## 3. Method

In this section, we briefly discuss the methodologies used in each individual study in our meta-synthesis (see Table 1). We then discuss our process for identifying common themes across the studies.

**Table 1**. Methodologies and study descriptions for data used in meta-analysis

| Study number | Methodology | Participants | Disciplinary focus | Year of data collection | Published work |
|---|---|---|---|---|---|
| Study 1 | Interview | Researchers (n=21) | Social sciences | 2017 | [18] |
| Study 2 | Interview | Researchers (n=19); Support professionals (n=3) | Multiple disciplines | 2017 | [19] |
| Study 3 | Interview | Support professionals (n=6) | Social sciences | 2016 | [6] |
| Study 4 | Survey | Researchers (n=1630); Support professionals (n=47) | Multiple disciplines | 2018 | [20, 21] |
| Study 5 | Observational study | Researchers (n=12) | Social sciences | 2019 | [22] |
| Study 6 | Use case analysis by support professionals | Use cases (n=100 use cases collected; n=25 support professionals) | Multiple disciplines | 2020 | [24] |

### 3.1. Methodologies for individual studies

#### 3.1.1. Study 1: Interviews with researchers

*Study and primary analysis description*

In Sun [18], twenty one face-to-face semi-structured interviews were conducted between August and December 2017. Participants were recruited through convenience sampling to identify the types of information needed for reusing quantitative social science data. Participants discussed their experiences with data reuse and performed observed searches in the Inter-university Consortium for Political and Social Research (ICPSR) data repository to find data meeting their needs. The interviews were audio-recorded and transcribed. Transcripts were analyzed using the open coding and axial coding techniques introduced by Corbin and Strauss [67].

*Sample description*

Twenty-one faculty members and research staff at two universities in Singapore (Nanyang Technological University and National University of Singapore) were recruited via email invitation. All participants were social scientists using quantitative research methods, had interest in or experience with reusing quantitative research datasets, and self-reported working in different domains (sociology (n=11), communication studies (n=6), international studies (n=2) and business (n=2)).

*Secondary analysis presented in this paper*

The previous analysis of the interview study focused on identifying the type of information researchers used to search, browse, and assess the reusability of research data. Here, we focus on reporting details about their data discovery behaviors and compare these with the results from other studies.

**3.1.2. Study 2: Interviews with researchers and support specialists**

*Study and primary analysis description*

As further described in Gregory et al. [19], twenty-two one-hour, semi-structured interviews were conducted between October and December 2017, either virtually or in person. Participants (n=19) were recruited via email from a pool of 186 individuals who had visited the DataSearch portal (now Mendeley Data[10]) and had indicated willingness to provide feedback. An additional three participants were recruited via purposive sampling. The interview protocol focused on eliciting information about participants' data needs, sources and strategies used for finding data, and their criteria for evaluating data which they discover. Audio-recordings were made of all interviews. Detailed summaries, including quotations with timepoints, were made following a protocol for every interview. Summaries were thematically coded and analyzed using QDA Miner Lite.

*Sample description*

Nineteen participants identified as being researchers; three participants worked as support specialists. Respondents were employed in 12 countries; the most frequently represented were the United States (n=6) and the Netherlands (n=3). Participants self-identified their broad disciplinary area and computer science (n=3) and information science (n=3) were the most common. Participants also had varying levels of professional experience: early career (0–5 years, n=5), mid-career (6–15 years, n=10), experienced (16 + years, n=6) and retired (n=1).

*Secondary analysis presented in this paper*

The interview study was previously analyzed primarily from the perspective of researchers. Here, we report further details about researchers' data needs which were not previously published, analyze the perspectives of support specialists, and compare these with the results from later studies.

---

[10] https://data.mendeley.com/

### 3.1.3. Study 3: Interviews with support specialists

*Study and primary analysis description*

As described in Friedrich [6], six support specialists from the German GESIS data archive for the social sciences were interviewed regarding their interactions and experiences with users of the data archive in June and July 2016. The goal of the initial study was to find out more about data users' information seeking behavior. Before surveying the users of the GESIS data catalog directly, Friedrich [6] conducted these interviews with support specialists. The interviewees served as a cumulation point for users' practices and problems when looking for data. That way, the interview data helped inform the questionnaire development for the survey. Following constructivist grounded theory methodology, the interviewees were sampled by initial and theoretical sampling, and the interviews were designed as unstructured conversations exploring eight topics: the data users' educational/professional background; the users' research experience and data literacy; information sources and channels; the roles of intermediaries and information technology; goals, needs and purposes of users; requirements when looking for data; the role of documentation; barriers and problems when looking for data. The shortest interview lasted about 52 minutes, the longest 1 hour and 36 minutes. Audio recordings and transcripts were made of all interviews. The transcripts were analyzed according to constructivist grounded theory methodology, using the software atlas.ti.

*Sample description*

All six interviewees were employees of the same large social science data archive. Two participants were help desk staff who received requests via the general helpdesk line. The other four were specialized experts for data from one or more complex large-scale social science surveys. The participants had been working as data support staff for different periods of time. The most junior staff member had been working in their position for 3 years, the most senior staff for 25 years.

*Secondary analysis presented in this paper*

As indicated in Friedrich [6], the interview data that had been collected for the purpose of designing a survey questionnaire for users of the data catalog has potential for further analyses regarding other research questions. In particular, the data does not present a direct picture of data user behavior but rather reveals the support specialists' perspective on users' data needs, data seeking practices and possible problems. In combination and comparison with other data, this data on the perspective of support specialists has the potential of revealing differences in perception of data services. Re-analyzing the data in that way may point us to challenges, problems and needs of users that support specialists see differently or not at all.

### 3.1.4. Study 4: Survey with researchers and support specialists

*Study and primary analysis description*

As described in Gregory et al. [21], a global survey was conducted to investigate practices of data discovery and reuse in September-October 2018. Respondents were recruited by emails sent to a random sample of 150,000 authors who published an article indexed in the Scopus literature database between 2015-2018. An additional 40 participants were recruited by posting to discussion lists in the library and research data management communities for a total sample of 1677 complete responses. The survey consisted of a maximum of 28 individual items, with 9 items allowing for multiple responses. Responses were recorded anonymously and were analyzed using R [68], the data from the survey are openly available at Gregory [20]. This survey was descriptive in nature; responses only describe the practices of people completing the survey.

*Sample description*

A total of 1630 researchers and 47 research support specialists completed the survey. Both the majority of researchers (68.4%) and support specialists (74.47%) reported working at a university or college, followed by those working at a research institution (17.18%, 14.89%). The majority of both researchers (40.18%) and support specialists (46.81%) have been active professionally for 6-15 years. The second largest group of researchers had 16-30 years of experience (30.25%), while the second largest group of support specialists had 0-5 years of experience (27.66%). The most selected disciplinary domains of researchers were engineering and technology, biological sciences, medicine and social sciences. Among support specialists, the most selected domains were information science, environmental science, multidisciplinary research and social science.

*Secondary analysis presented in this paper*

Survey results regarding researchers were published in Gregory et al. [21]. Here we analyze the survey data collected from support specialists and compare it to the data from researchers. New visualizations comparing both the perspectives of support specialists and researchers were created using Tableau. This allows a comparison which was not yet performed in previous research.

### 3.1.5. Study 5: Observations of researchers

*Study and primary analysis description*

An observational study described in Krämer et al. [22] was conducted with 12 social scientists recruited through convenience sampling to learn about their data-seeking behaviors. It consisted of an initial survey, an observational period in which screen recordings were collected while participants performed a simulated data search task, and a follow-up guided interview. The audio material was transcribed. Screen captures were transcribed to show interactions with the computer in an abstract way, i.e., using web search, university library web page, etc. The interactions were also analyzed separately for patterns. The merged transcriptions were analyzed using affinity diagrams.

*Sample description*

At the time of the study, eleven participants were affiliated to a university and one to a public research facility. Four of the participants were professors, three were postdocs, two were PhD students, and three were master's students. Their primary research areas were sociology (n=6), political science (n=4), economic sociology (n=1), and media management (n=1).

*Secondary analysis presented in this paper*

Krämer et al. [22] focuses on giving an overview of the observed behaviors giving a total of nine key findings. For the secondary analysis, we offer more details on interconnectedness of web search and literature search in particular. These details come from re-examination of the source materials, in particular the transcripts.

### 3.1.6. Study 6: Use case analysis and ranking by support specialists

*Study and primary analysis description*

As described in [24][11], the GOFair Data Discovery Implementation Network[12] collected over 100 use cases for data discovery by polling researchers and support specialists and by consolidating existing lists, i.e., [23]. These use cases were then clustered around common themes and requirements. A survey tool administered during the GOFAIR meeting (February 3rd, 2021) was used to gain consensus about the identified themes. Meeting attendees were asked to answer a demographic survey (position, domain and geographical location) and to rank the two most and least 'relevant' use case clusters.

*Sample description*

25 out of approximately 50 meeting attendees completed the survey. Most respondents were from mainland Europe (n=17), with three additional participants from the UK (n=3). Of those mainland participants, most were located in Germany (n=8), Scandinavian countries (n=4) and the Netherlands (n=3). Three attendees were from North America (2 US, 1 Canada), and one from South America (Brazil). The majority of respondents identified themselves as infrastructure providers (n=17), with eight respondents indicating that they produce data themselves. Most respondents were involved in the life sciences (n=8), general science (n=7), or natural sciences (n=5). Fewer worked in social science (n=3) and computer science.

*Secondary analysis presented in this paper*

Mathiak et al. 2023 [24] only includes a very short analysis on how the ranking of use contrasts with the needs found in user studies. This aspect is elaborated much further in this work, by re-examining the survey data and differentiating different aspects, such as web search, literature search, social networks.

### 3.2 Methodology for meta-synthesis

As authors of the above studies, we are all deeply familiar with the methodologies and resulting data. To synthesize the data from across all studies, we conducted four phases of analysis. First, we engaged in an exploratory phase, where we each took our individual expertise as a starting point. We each re-examined the data from our own studies above, with a focus on supporting professionals' data discovery and reuse practices and attitudes and our research aims. We then summarized our findings textually and graphically and shared them with each other before collectively identifying themes in our individual analysis. Finally, we compared these individual themes with each other and identified five cross-cutting themes which we use to present our findings below. In our conclusion we discuss these findings using the typology of support specialists that we have introduced in the literature review.

The aim of our meta-synthesis is to compare perspectives of support specialists and researchers across disciplines. We recognize, however, that data, research norms, and available infrastructures in different disciplinary communities impact data discovery. Three of our studies focus on social science data; three others include data from individuals in multiple disciplines, including those in the social sciences (Table 1). In our analysis, we therefore also compare the results from our social science studies to multidisciplinary studies.

---

[11] Note to the editor and reviewers: This paper has been accepted with revisions at the CODATA Data Science Journal. We are very optimistic that we will be able to include the citation in the final draft of this paper. Otherwise, we can point to the pre-print, which is already available.

[12] https://www.go-fair.org/implementation-networks/overview/discovery

## 4. Findings

Our findings are arranged in five thematic clusters: challenges and problems in data seeking; data needs and purposes; the role of web search in data discovery; the role of academic literature; and the role of social connections. Within each cluster, we present findings from both the researcher perspective and the support specialist perspective.

### 4.1 Challenges and problems in discovering data

#### 4.1.1. Researcher perspective

Seventy-three percent of researchers responding to the global survey (Study 4) indicated that finding data is sometimes a challenging task, while 19% of respondents believed it to be difficult. Survey respondents identified specific challenges, including the inability to download or access data (68% of respondents) and the distributed nature of data across different locations (49%). Other identified challenges included a lack of adequate data search systems, a lack of data search literacy, or, as found in the semi-structured interviews reported in (Study 2), a lack of digital and discoverable data.

At the same time as reporting these challenges, interviewees in Study 2 felt that they were ultimately satisfied with their discovery strategies, although they recognized that they need to modify their expectations or employ multiple strategies (including consulting their social networks and personal connections) to locate the data which they use. Researchers also reported seeking help from colleagues to design efficient queries; others educate their own colleagues about sources and strategies to use when searching for data online. These more 'social' strategies, which are discussed further in Section 4.5, also have challenges. Study 2, e.g., found that relying on social networks alone can lead to the creation of 'filter bubbles' where researchers may only find data available from within their own circle of peers.

In the pre-observation questionnaire used in Study 5, it was found that finding data is perceived to be difficult in comparison to literature search, which is usually perceived to be of low or medium difficulty. This is similar to results in [69]. One of the challenges identified in the same questionnaire is low data search literacy. During the interviews in Study 5 more challenges emerged: there were not enough links between datasets and relevant other information sources, i.e., other datasets, literature and documentation. Limited accessibility was also mentioned as a challenge to a successful dataset search.

#### 4.1.2. Support specialist perspective

In Study 3, the interviewed support specialists reported that users requested help for both exploratory search and known-item searches, particularly in locating suitable data sources. Users also requested help when they had problems accessing data. Access-related problems included challenges related to data location and retrieval, cost of access, and restricted access.

The interviewees also discussed problems users face when they have found relevant data but cannot make use of them. These problems fall into three categories: problems with erroneous data, when users find potential errors in the data; problems with documentation, when users need assistance in finding or understanding documentation; and problems with data literacy skills, when users lack the necessary skills to actually use data they have found.

In Study 6, research support specialists were asked to prioritize clusters of data discovery use cases to identify the most urgent areas for future work. The cluster of use cases related to *metadata/metadata quality* was ranked as being most important, followed by the clusters *data citation* and *convenience*, the last of which refers to improving the user experience of data search. Participants were also asked which clusters they felt were least important in data discovery. Here, clusters related to lesser known approaches, such as search within data, linking with persons, and machine discoverability were seen as having low importance. What is somewhat contradictory is that while metadata quality was rated as highly important, the documentation cluster was often rated as least important.

**4.2 Data discovery needs**

**4.2.1. Researcher perspective**

Both Studies 2 and 4 highlight that researchers need multiple types of data, often with very specific resolutions and parameters, across phases of work. In Study 4, for example, researchers classified the type of data which they need using a categorization system from the National Science Board [70]. While the majority of researcher respondents selected observational data, approximately half also selected needing more than one type of data. In Study 4, researchers also indicated that they use data for a variety of purposes, ranging from using data as the basis for a new study to using data for teaching. Figure 1, discussed further in Section 4.2.2, presents the distribution of purposes selected by respondents according to their professional role.

Study 1 identifies ten types of uses social scientists have for reusing data. These uses, which are uses that will be specifically reported in publications, were classified into three categories: to support the conceptualization of the study design; to complement the main study, or to serve as the primary data of the main study. The interviews in Study 2, however, find that data reuse can also be conceptual and not be reported in the final paper, even if data have informed the study design and development.

In Study 5, we observed on multiple occasions that social scientists decided to refine their original research question based on the data they could find. They either shifted their topic or research question, or more often, made other search refinements, such as narrowing a general search query into more specific terms.

**4.2.2. Support specialist perspective**

The data needs of support specialists are often the data needs of their clients. One of the interviewees in Study 2, e.g., searched for data in the literature about particular technologies such as intravascular ultrasounds to pass on to her clients. Finding data and associated literature is just one way in which support specialists work to meet the data needs of their clients. Support specialists also report engaging in educational activities such as teaching people how to discover and evaluate data or how to curate and manage their own data (Study 4). Study 3 shows that support specialists get requests from users who need data for a broad array of purposes. The purposes include not only various research-related tasks, but also using data for professional qualification and in journalism.

Support specialists also need data for their own research and projects. Forty percent of support respondents in Study 4 (n=19) reported needing data both for their own research and to support others, while fewer individuals (n=2) need data solely for their own research.

There are some differences between how researchers and support specialists use - or view the use of - secondary data by researchers (Figure 1). In Study 4, respondents were asked about their purposes for seeking data. Researchers were asked "Why do you use or need secondary data?", while support specialists were asked "Why do you or the people whom you support use secondary data?"

**Fig. 1** Purposes for using secondary data by role: researchers (n=1630) and support specialists (n=47). Multiple responses possible.

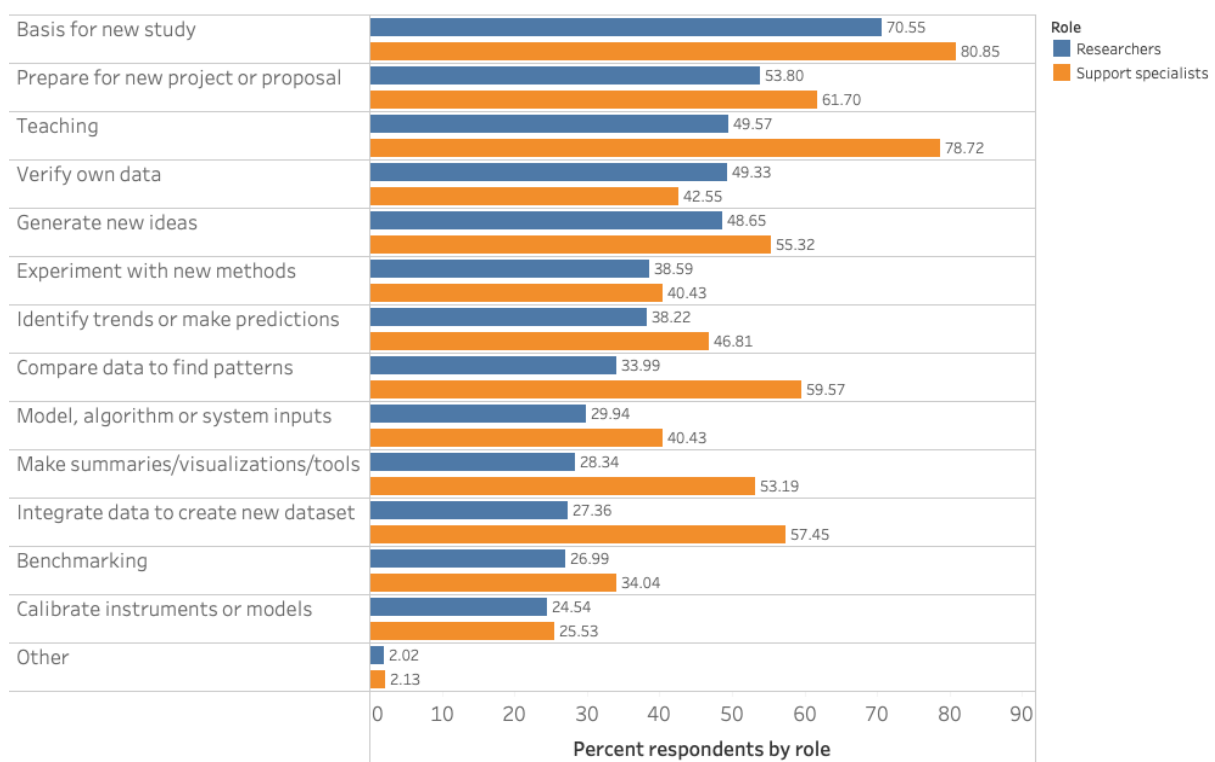

Approximately 50% of researcher respondents use data for teaching, while nearly 80% of support specialist respondents selected this use. In a separate question, support specialists indicated that they help others through teaching; the high frequency of using data for teaching indicated in Figure 1 mirrors this practice. Similarly, support specialists selected using data to make summarizations or visualizations at a higher percentage than researchers; this could be attributable to their own teaching practices also.

Support specialists also indicated that they integrate data (or believe that researchers integrate data) at higher percentages than did researchers. This could reflect the fact that data integration can be a key step in curatorial workflows for certain data, e.g., long-running survey data in the social sciences. It could also indicate a potential mismatch between how support specialists view the activities of researchers, as data integration is often discussed in both the literature and efforts around FAIR data, which is perhaps reflected in the answers of support specialists.

Support specialists in Study 3 described the data needs of users seeking survey data, as presented in support requests. This data was used to identify six categories of data needs, as presented in Table 2 below. Interviewees in Study 3 reported that requests with only few specifics or very general requests require more support than very specific requests.

**Table 2**. Data needs of users seeking survey data

| Data need | Example requests |
| --- | --- |
| General need for any kind of data | "What data do you have?" |
| Need for new research ideas | "Do you have data on any exciting topics? |
| Need for data on a specific topic | "Which new or novel data do you have on this topic? |
| Need for known data | "Someone recommended this data to me; do you have it?" <br> "Do you have an updated version of this data?" |
| Need for specific variables in known data | "I know the data I need, but I can't find the variable I am looking for." |
| Need for data with specific features | "I need longitudinal data / linked data / comparative data." |

**4.3 The role of web search in data discovery**

**4.3.1. Researcher perspective**

Across disciplines, general web search engines, such as Google or Bing, are important sources for researchers looking for secondary data. 89% of researchers in Study 4 report using general web search engines occasionally or often to locate data for reuse. They use web search both to conduct known-item searches, to locate data repositories, and to search for new datasets, a finding which was also supported in the interviews conducted in Study 2.

Study 2 found that the researchers' success of using web search engines to locate needed data was mixed. This mirrors the findings from Study 4, where 54% of the responses from researchers indicate that searching the web to find data is sometimes successful and sometimes not; 39% of researcher responses indicate that web searches are either successful or very successful.

In Study 1, most social scientists directly searched for known datasets using general web search engines to identify the digital location of the needed data. In total, the 21 interviewees reported three types of online sources to obtain publicly available datasets: 1) the official website of a particular study, such as the Health Information National Trend Survey (n=13); 2) databases hosted by data collection organizations, such as the World Health Organization (n=9), and 3) data repositories such as ICPSR (n=3).

Study 5, the observation study, found that web search was used extensively for a number of purposes. Participants often performed known item searches, i.e., by searching for the names of datasets that were found in the literature, but they also used web searches to find additional information on datasets, such as documentation, study websites, or related literature. Participants used web search as a quick and convenient way to chain together different resources, which would otherwise not be connected.

### 4.3.2. Support specialist perspective

There is an overall lack of information in our studies, and in the available literature, which documents how support specialists locate data by searching the web. Study 4 suggests that researchers may rely on general web search engines more heavily than support specialists, as more respondents who are researchers (59%) reported *often* making use of web search engines than did support specialists (40%). This difference becomes much smaller, however, when looking at the percentages for respondents who selected *never* using search engines to find data (11% of researchers and 13% of support specialists).

There was also less variety in how support specialists reported their success using web search engines to find data, with 80% of responses from support specialists indicating that they are sometimes successful/sometimes not successful in their data searches with general web search engines.

Support specialists interviewed in Study 3 indicated that researchers use web searches in particular for finding data from studies that they already know (known-item searches). They also indicated that users expect data repositories to be as easy to use as general web search engines, which they often are not.

In contrast, support specialists in Study 6 did not recognize or prioritize the role of web search in searching for data. None of the over 200 use cases in this study refer to web search and therefore had no priority assigned to them. While absence of proof is not proof of absence, web search does not seem to be the first thing that came to the minds of the many support specialists who compiled the list of use cases in the study. This is different from what we found for our next other two topics: using literature and social networks, which are both well-represented in Study 6.

### 4.4 Data discovery source: Literature and data citation

### 4.4.1. Researcher perspective

Study 2, Study 4 and Study 5 document the importance of the academic literature in researchers' practices of discovering data. Interviewees in Study 2 reported following citations to data, plucking data, including device specifications, from figures, tables, graphs and the supplementary information attached to articles. Literature search was highly prevalent in all observations of data search made in Study 5, with one person stating that literature search "enables data search".

Seventy-five percent of researcher respondents to the survey in Study 4 reported *often* using the literature to find data, with 19% of them stating that they occasionally did so. Researchers make use of the literature in many ways, some of which are different from how support specialists do so (Figure 2). Both support specialists and researchers use the academic literature to find data by following citations. More researchers turn to the literature with the explicit goal of finding data from the literature than do support staff.

**Fig. 2** How respondents use the literature to discover data by role: researchers (n= 1531) and support specialists (n=42). Question only asked to respondents who report using literature as a data discovery source. Multiple responses possible.

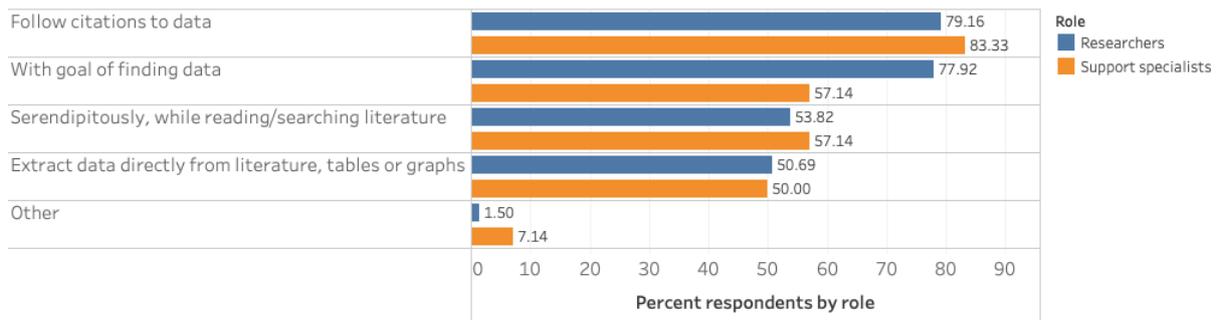

Study 5 reports that most of the participants used literature for their data search. Some used it to assist with the search, e.g., for relevance assessment, or to see how someone else did an analysis. Other participants conducted their whole dataset search through literature search, seeing literature search as being a prerequisite for searching for datasets. Similarly in Study 1, one social scientist's normal practice involves discovering reusable variables in datasets by first searching for relevant literature.

### 4.4.2. Support specialist perspective

Study 4 suggests that support specialists make more use of a diversity of sources to locate data than do researchers, with respondents selecting that they *often* use the literature (42%), search engines, domain repositories and governmental sources at roughly the same percentage.

Support specialists tend to view data discovery and literature discovery as separate and distinct practices, whereas more researchers turn to the literature with the explicit goal of finding data from the literature than do support staff (Figure 2). In Study 4, 49% of support specialists (n=23) stated that their process for finding data and for finding literature is always different, while less than 20% of researchers responding to this question say that their processes are always different (Figure 3).

**Fig. 3** Responses to the question "Do you discover data differently than the literature?" by role: researchers (n=1630) and support specialists (n=47).

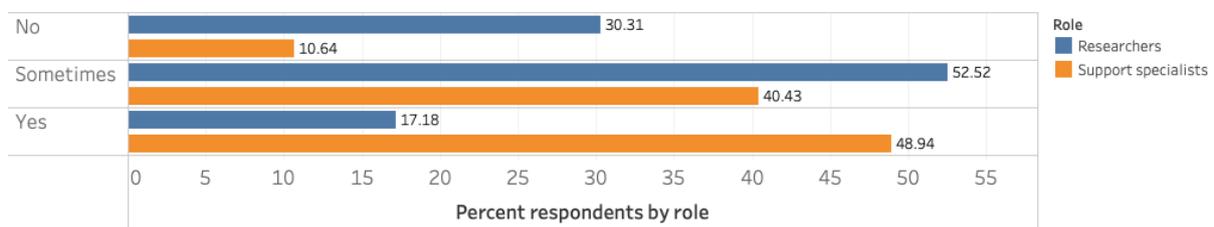

In Study 3, the interviewed support specialists reported that research papers are one of the main sources for learning about existing data for data archive users. One support specialist also reported observing a frequent use of DOI references to enter records in the data catalog, which suggests that users are citation chaining from papers. In particular for students or young researchers, textbooks were seen to be another important source for discovering data.

Of all the data discovery use cases collected by support specialists in Study 6, data citation and more generally, the connection between literature and datasets, had the greatest number of use cases. However, when it came to prioritization, some ranked it as high priority, while others explicitly ranked it as low priority. (Participants could only pick two of each).

### 4.5 Data discovery via social connections

### 4.5.1. Researcher perspective

In Study 4, researchers report often (31%) or occasionally (54%) using personal networks to discover data. The ways in which researchers and support specialists make use of social connections to *discover* data has similarities (i.e., both groups report relying on personal conversations) and differences, namely in that researchers report attending conferences as a way to discover data more and support specialists report making use of disciplinary forums (Figure 4).[13]

**Fig. 4** How respondents use social connections in data discovery. Researchers: discovering (n=3192), accessing (n=3497), sensemaking (n=2949) and support specialists: discovering (n=119), accessing (n=92), sensemaking (n=82).

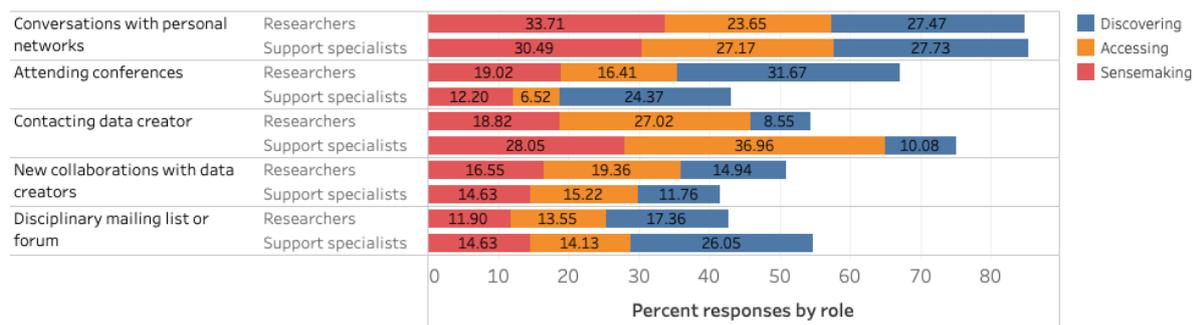

There are also similarities and differences in how the two groups make use of social connections to *access* and *make sense* of data. Researchers selected forming new collaborations with data creators and attending conferences slightly more often than did support specialists. Support specialists selected contacting data creators as a way to access and make sense of data more frequently.

Study 1 also revealed that most of the interviewed social science researchers mainly used personal networks to *discover* and *access* data. They reported two types of reused data: publicly available data downloadable from the web and non-publicly available data obtained from research collaborators. For the former, most interviewees reported that they could not recall spending particular effort in data discovery. Rather, their knowledge of the existence of the data was accumulated incidentally from past research experiences (e.g., training, hearsay from colleagues). For the latter type, interviewees described obtaining access to the data through collaborating with data collectors (e.g., PhD advisors, colleagues, or non-research organizations that pay researchers for data analysis).

In Study 5, participants explained in the post-observation interview that they would get information about datasets from colleagues. During the observation, when given the option to communicate with colleagues, one participant made a phone call, and others wrote emails or said they would write an email. Supervisors played a special role for younger scientists, in that they recommended both the datasets as well as guided on data analysis. It seems that secondary analysis is often teamwork with specialized roles among team members, such as defining research questions and analyzing data.

### 4.5.2. Support specialist perspective

---

[13] Responses also indicate envisioned behavior of researchers.

Personal networks are also important for support specialists, as indicated in Figure 4 above. Study 3 revealed that community involvement plays an important role in finding data. According to the requests that data support specialists receive, researchers request data that they have been pointed to, for example by supervisors. Others exchange datasets with peers, even data with access restriction (data dealing). The interviews also revealed that support specialists themselves make use of a larger data-related community when catering to users' needs. These data communities include, for instance, principal investigators who have collected the data and work closely with data professionals (data managers, data curators, data archivists, data librarians and others) to prepare the data for archiving, publication or reuse. The interviews also suggest that data reusers themselves are part of these communities, for example, when they are invited to detect and report errors in datasets or make other suggestions for improvements or enhancements.

In Study 6, the use case cluster regarding social connections was one of the smallest. This cluster was not seen as relevant by the support specialists.

## 5. Synthesis and Analysis of the Findings

In this paper, we conducted a meta-synthesis and re-analysis of six of our own studies on data discovery, with the aim of bringing together the perspectives of both researchers and research data support specialists. In doing so, we identified both points of alignment and difference in the two perspectives in all five of the themes which we identified in our analysis (Table 3). We discuss these points of alignment and difference here, before relating them to possibilities of future work for support specialists in Section 7, according to the typology of support specialists identified in Section 2. We also differentiate in our discussion in this section between multidisciplinary studies and those describing practices in the social sciences. In both Table 3 and in the text we indicate social science studies using an asterisk, i.e. Study 1*, Study 3*, and Study 5*.

While we find high alignment on the challenges and data needs between researchers and support specialists, there are three areas of lower alignment which emerge in our analysis: web search, use of literature and social connections.

**Table 3**. Comparison between the findings of researchers and support specialists (cf. Section 4) and their alignment. Numbers in parentheses refer to study numbers; social science studies are indicated with an asterisk.

| Theme | Findings: Researchers' Perspective | Findings: Specialists' Perspective | Alignment |
|---|---|---|---|
| Challenges | data discovery is challenging (2, 4, 5*) <br> lack of search tools (2, 4, 5*) <br> lack of data (seeking) literacy (4, 5*) <br> data accessibility (4,5*) <br> modifying strategy to fit the data (2,5*) <br> lack of links (5*) | researchers need assistance (3*) <br> see a challenge to locate and access data (3*) <br> data literacy of researchers too low (3*) <br> erroneous data (3*) <br> priorities: metadata, data citation, ease of use (6) | High |

|  |  | non-priorities: documentation, links to persons, machine discoverability (6) |  |
| --- | --- | --- | --- |
| Data needs | need multiple data types (2,4) multiple different purposes for data (1*) sometimes very specific and complex needs (2,4) | specialists inherit researchers needs (2,3*,4) less likely to conduct their own research (4) more likely to use data for teaching (4) specialists are perhaps more likely to do data integration/curation (4) | High, with slightly different emphasis |
| Web search | web search is a highly used data discovery tool (1*,2,4,5*) used for a variety of purposes (5*) | do not use web search as much (4) do not recognize web search as part of data discovery (6) are aware that web search is important for known-item searches and that data repositories lack equal searchability (3*) | Low, this seems to be underestimated by support specialists |
| Use of literature | very important part of data discovery (1*,2,4,5*) "Enables data discovery" (5*) | use literature differently than researchers, but also heavily (4) aware that researchers discover data by citation chaining from papers (3*) literature is a priority for some specialists, but not for others (6) see literature search as being more separated from data discovery (4) | Medium, with slightly different emphasis |
| Use of social connections | important part of data discovery (1*,2,4,5*) used for finding and making sense of data (4,5*) | rely less on social networks for data discovery than researcher (4) expect to contact data creators directly (4, 3*) is not seen as a priority (6) | Medium, some differences |

### 5.1 High alignment between researchers and support specialists

*Challenges*

Data discovery is perceived as challenging by researchers and support specialists alike. Both Study 4, a multidisciplinary study and Study 5*, a social sciences study, found that challenges in data seeking as reported by researchers are mostly about lack of data search skill, lack of data literacy, and lack of access to data. Study 3* reveals that support specialists who provide data services to researchers are aware of these challenges and work to support these challenges through training and consultancy efforts. This is supported in [39].

In Study 2, 4 and 5*, researchers mention technical challenges to data discovery, such as a lack of suitable search tools and links between research data and other relevant sources of information, in particular the academic literature. Support staff in our studies don't see priorities in this area. Instead, they observed (Study 3*) or self-assessed (Study 6) that the quality of data or data documentations is an important factor supporting data seeking behaviors. This reflects the fact that support specialist work includes tasks related to metadata curation and improving metadata and data standardization [4, 61, 71].

Though in our findings, data being distributed across different locations is reported as a challenge by researchers (Study 4) but neglected by support specialists, it is a challenge recognized by support specialists and information professionals in other studies, which describe the development of metadata standards and tools for federated dataset search to support data discoverability (e.g., [60]).

*Data needs*

Studies 1-5, including multidisciplinary and social science studies, provide evidence that researchers have multiple, evolving data needs, ranging from the more general to the more specific. Not all of these uses may be cited in a research publication (Study 1*, Study 2), some data use happens outside academic research (Study 3*). The data needs of support specialists commonly reflect the data needs of researchers in both social sciences and multidisciplinary studies (Studies 2, 4, 5*). The interviewed support specialists from Study 3* proved to be aware of both very general as well as very specific needs and purposes for data use. Support specialists sometimes need data for their own research (Study 4), albeit to a lesser extent than researchers. Studies 1-4 also highlight that both support specialists and researchers need data to use in their own teaching.

Many support specialists' own data needs are service-oriented and are tied to the support work that they do (i.e., teaching, summarizing, or visualizing data). Study 4 suggests that support specialists' perceptions and researchers' actual data needs may not always be in alignment, and that support specialists are perhaps more invested in pooling or integrating datasets. This finding may reflect the curatorial nature of some support specialists' work, i.e., integrating and maintaining longitudinal survey data, as found in prior studies (e.g., [4]).

**5.2 Medium alignment between researchers and support specialists**

*Use of social connections*

With exception of Study 6, all studies show that social connections are viewed as important by both researchers and support specialists. The ways of connecting may be different as Study 4 suggests. For researchers across disciplines, social connections are an important means to find data (Study 1*, 2, 4 and 5*) and are of value in reusing and making sense of data (Study 5*). This reliance is in line with many other studies on data discovery practices and data reuse [13, 28, 72, 73].

 Study 1* suggests that the relevance of social connections may depend on the accessibility of the data; finding open datasets may not require social interactions as much as finding access-restricted datasets. Study 3* shows support specialists confirm this finding and also describes the composition of these social networks, which include data producers, primary researchers, and support specialists. For support specialists, these connections are important to prepare data as well as to assist researchers in finding and reusing data.

Despite this awareness, social connections are not defined as a formal way of data discovery that deserves attention from the support specialist community (Study 6). This contrasts the fact that support work related to data discovery is itself a collaborative process that involves many personal exchanges (Study 2, [42]).

*Use of literature*

Another point of medium alignment of perspectives is the use of academic literature. All six studies - multidisciplinary and social science studies - show in various ways that academic literature is important for researchers to data discovery. The degree of importance, however, seems to be less pronounced among data support specialists (Study 4 and 6), although they do seem to recognize that this is common practice among researchers (Study 3*). Our findings also indicate that support specialists use literature less themselves as an avenue for finding data and perhaps see data discovery as a practice that is more distinct from literature discovery. Nevertheless, there are increasing efforts in linking research data to papers to enhance data discoverability such as Scholix [51] and knowledge graphs [49, 50, 52, 74].

### 5.3 Low alignment between researchers and support specialists

*Web search*

As shown in Studies 1*, 2, 4, and 5*, in both multidisciplinary and social science studies, researchers use general web search engines extensively to discover data, as they search the web to find data repositories, conduct known-item searches and to find data directly. This is also found in prior studies [34].

Study 3* shows that support specialists are aware of this practice. They also concede that the retrieval tools at data repositories do not meet users' expectations regarding searchability. There also appear to be differences in perspectives. Study 4 suggests that support specialists may rely less on web search but rather spread their efforts across sources. Study 6 likewise suggests that support specialists may not see web search as being important enough to consider when thinking about supporting researchers' data discovery practices, as it was not mentioned in the use case ranking.

## 6. Discussion

### 6.1 Heterogeneity of the studies

Putting all these different findings together is challenging given their different methods, the different demographics of the subjects, and different levels of abstraction. The studies in their variety of methodologies do not support representativity, but rather provide in-depth data from a variety of perspectives.

There may also be subtly different understandings of the data discovery process in play across the studies. It is not always clear whether this process includes data reuse practices or data search behaviors specifically. Given these vagaries, we limit our findings to general observations, such as high alignment and low alignment of certain topics and excluded any findings we could not find consensus on.

As mentioned in the literature review, there are even difficulties in defining what a "data support specialist" is. This term could refer to a librarian, a data archivist with a domain background, a metadata expert, or an IT specialist. We address this difficulty by creating a typology of work foci in the literature review. This typology, which includes people-oriented, metadata-oriented, and infrastructure-oriented work, is used to make the recommendations regarding future work in Section 7. However, it should be noted that retroactively applying these categories to our own works has proven difficult and, in practice, the categories often overlap.

Research data management is a quickly evolving topic, and the studies we analyze cover multiple years, in which much has changed. The FAIR data principles [75] evolved from an idea in 2016 to a widespread movement in 2022. The field of data discovery has evolved from being a specialized subcategory within data reuse and information retrieval research to becoming a distinct topic with dedicated support, i.e., from associations such as the Research Data Alliance. Despite these wider changes, we still see consistency in our results across time. It could be posited that social connections have become less important, and literature and web search have become more important, but the difficulties in combining the different studies methodologically do not allow us to make statements of the necessary accuracy.

### 6.2 Main insights from the meta-synthesis

Many of the themes we discovered in this work are consistent with those from our previous research studies, and align with other studies which have found that researchers, in particular, rely on social connections, literature, and web search to locate data, even, as seen in Study 1, when existing disciplinary data infrastructures are available [28, 34, 72, 73]. These findings further support the idea that these are key strategies for researchers in discovering data.

The recent review by Chapman et al. [76] stresses the importance of metadata quality on data discovery, which echoes the perspectives of support specialists which we have identified. However, they do not compare the perspectives of the groups of support specialists and researchers, which is unique to our analysis.

In our meta-synthesis, we not only compare the practices of support specialists to those of researchers, as highlighted in Table 3, but we also pay close attention to findings from studies focusing on support specialists that have not received as much attention or which have not been previously published. Of these findings, three are of particular interest.

While the use of social connections to find data has been increasingly well-documented in previous research, our meta-synthesis, specifically results from Study 3*, highlights the active involvement of support specialists in research and 'data communities.' Support specialists leverage their social connections to assist researchers in finding data; this provides a new perspective on the utilization of social connections in data discovery and emphasizes the crucial role that support specialists play in fostering data discovery within research communities. Data communities, as we use the term here, consist not only of principal investigators who have collected data, but also of various data professionals such as data managers, curators, archivists, librarians and others. This suggests that the distinction between 'research' and 'support' roles may become less clear as these two professions are brought together in these communities, as also suggested in [77]. Additionally, our findings from Study 4 suggest a lack of distinct boundaries between research and support roles. Support specialists not only support others in finding data, but they also need data for their own research and projects. This highlights the interdependence and overlap of the roles of researchers and support specialists in the data discovery process.

Another interesting point of difference from the perspective of support specialists is found regarding data integration. Study 4 shows that data integration is viewed differently by support specialists than by researchers, with more support specialists reporting this type of data use. We hypothesize that this could be because integrating data is an important aspect of curatorial workflows for support specialists.

A third point of distinction from the perspective of support specialists is the separation between data discovery and literature discovery. Figure 3 shows that support specialists across disciplines may tend to view data discovery as a distinct practice in comparison to literature search, while researchers tend to see them as more interconnected. This may be explained with the different roles both groups find themselves in. Researchers tend to be highly specialized within their domains. Literature is important to them to gauge the impact of their work, including working with data, within their community. Research support specialists, on the other hand, often search data as part of a delegated task, without the need to understand the bigger picture of the research. They have a more hands-on experience with data management, curation, and preservation, which could lead them to view data discovery as a separate practice that requires specific skills and expertise.

**6.3. Disciplinary differences and application of results**

While our research aims and study design necessitate high level comparisons and statements, we are also aware that disciplinary and community-specific data practices exist, both for support specialists and researchers. As a way of addressing and calling attention to some of these differences - as well as to identify similarities, which can be useful when designing tools and services - we highlight results from social science and multidisciplinary studies in our analysis of the findings (i.e., Section 5). Overall, we find much agreement between the results from both social science and multidisciplinary work, particularly in our high-level characterization of the use of literature and social connections. While there may be similarities at the general level of abstraction which is suitable in a meta-analysis such as this, our findings also point to the specific and local nature of data needs and practices, i.e., in both needing and supporting the location of particular survey variables or longitudinal studies in the social sciences (Table 2). While we pay special attention to the social sciences in this paper, we believe that our higher-level findings may be applicable to other disciplines, or that they may provide a starting point and inspiration for more local interrogation of the practices of researchers and support specialists.

**7. Conclusion**

**7.1 Recommendations for different areas of support work**

The alignments and differences we identify offer a window into thinking about how the work of support specialists could better support the data discovery practices of researchers. We conclude by thinking about how the different types of support work identified in Section 2.2 could address these points of difference, both by identifying existing best practices and by making suggestions for improvement.

**People-oriented type of support work**, especially in university libraries, may consider fully incorporating data literacy skill training and consultation (e.g., explicitly teaching the skill of searching data and evaluating results) into the existing research support toolkits. Data literacy is a complex concept; such training may range from "where to search for what data" to "advanced skills in data analysis". We see an opportunity for support specialists to provide different levels of data literacy training for researchers at varying stages of their research process. This may require support staff working with different disciplines to develop targeted training content.

On a similar note, we suggest that libraries offer training for students to find research data in the same way as they are already doing for students to find literature, which would be an extension of the traditional library-related reference support work to the new field of open data and RDM. Many university libraries are currently providing data literacy training as an *optional* provision on top of the more established information literacy skill training such as "how to use citation databases" or "how to search literature".

**Metadata-oriented support work** may need a general improvement with regard to the needs of researchers. While metadata quality is a problem, as seen in Study 3 where data seekers were not satisfied with documentation which has errors, another problem to address is to decide the right level of granularity of research contextual information included in data documentation. The usability of data documentation itself impacts the reusability of data.

Another promising strain of metadata-oriented support work is to continue expanding the scope and scale of the existing initiatives such as Scholix [51] to build links between research data and other research materials, i.e., literature, and research entities, e.g., researchers, research institutions. Our studies reveal that researchers need such links for discovering research data and evaluating data reusability (Study 1, 2, 4, 5), whereas support specialists don't assign link-building tasks as high priority (Study 6). This is a conspicuous discrepancy which has room to be addressed.

**Infrastructure-oriented support work** may fork into the development of search tools customized for different data reuse tasks, in line with the suggestions of Koesten at al. [28]. To do this, an in-depth understanding of data-centric reuse tasks is needed. We call for more user studies into researchers' data reuse behaviors, needs and requirements.

More user studies may also be needed to improve the findability of research data via general web search engines and to increase the usability of data repositories, which may be a topic for both metadata-oriented and infrastructure-oriented support work. It's not uncommon that researchers start with web search engines for searching and end up in repositories. While existing efforts focus on how to make research data more discoverable, i.e., through using Linked Open Data or the use of *schema.org* vocabularies, we also observe an equally relevant issue: retaining researchers after they land in a data repository.

Extant studies often use log analysis to investigate users' interaction with data repositories. We argue that more user studies with an eclectic mix of research methods (e.g., think aloud tasks or observational studies) are needed to inform the design of data repositories to improve both exploration of data and retention of data seekers.

Perhaps even more important is conducting studies that brings together researchers and supports specialists in co-design activities. We gather and compare practices and perspectives of both sides by re-analyzing data from our past work. Further research is needed to take this into account at the beginning of study design.

## 7.2 Future Work

Our work has shown that the alignment between users and research support specialists is not always perfect. We have argued that it is fruitful to examine differences to avoid building infrastructure that does not fit users' needs and have identified possible recommendations for bringing these two perspectives together.

We strongly believe that other studies directly comparing the perspectives of users and support specialists are needed. One way to address this it to design studies primarily designed for this purpose, rather than to use a re-analysis of existing data, as we do here.  Such studies could yield additional insight and more detailed answers to our research questions. We hope that our results sharpen awareness that these two perspectives are not necessarily the same.